\documentstyle[11pt,paspconf]{article}
\input tex.def
\input psfig.tex

\begin{document}

\title{Optical Spectroscopy of LINERs and Low-Luminosity Seyfert Nuclei
\altaffilmark{1}}
%\title{Optical Spectroscopy of LINERs and Low-Luminosity Seyfert Nuclei}

\author{Luis C. Ho}
\affil{Harvard-Smithsonian Center for Astrophysics, Cambridge, MA 02138}

\altaffiltext{1}{Invited review paper to appear in {\it The Physics of LINERs 
in View of Recent Observations}, ed. M.~Eracleous, A.~P.~Koratkar, L.~C.~Ho, 
\& C.~Leitherer (San Francisco: ASP)}

\begin{abstract}

An unprecedentedly large number of LINERs has been discovered in a recently 
completed optical spectroscopic survey of nearby galaxies, allowing several
statistical properties of the host galaxies and of the line-emitting 
regions to be examined reliably for the first time.  As a consequence of 
the many detections and some revised classifications, the detailed 
demographics of emission-line nuclei have been updated from those given in 
older surveys.  Consistent with previous studies, it is found that LINERs are 
extremely common in the present 
epoch, comprising approximately 1/3 of all galaxies with $B_T\,\leq\,12.5$ mag.
If all LINERs are nonstellar in origin, then they are the dominant constituents
of the active galactic nucleus population.  Many fundamental characteristics of 
LINERs closely resemble those of low-luminosity Seyfert nuclei, although 
several aspects of their narrow-line regions appear to differ in a systematic 
manner.  These differences could hold important clues to the key parameters 
controlling the ionization level in active nuclei.  Lastly, a substantial 
fraction of LINERs has been found to contain a broad-line region, yielding 
direct evidence, at least in these objects, of a physical link between LINERs 
and classical Seyfert 1 nuclei and QSOs.

\end{abstract}

\section{Introduction}

This workshop has repeatedly stressed the utility of observations in 
previously underused spectral regions, particularly the ultraviolet and 
X-rays, for understanding the physical nature of low-ionization nuclear 
emission-line regions (LINERs; Heckman 1980b).  Nevertheless, the amount of 
data on LINERs available from space-based instruments is still very small, 
limited by a combination of the long exposure times necessary to detect the 
faint signal in these sources and the usual oversubscription of spacecraft 
time.  Thus, statistical studies of LINERs, which require large quantities of 
data to be gathered in ``survey'' mode, still must rely on more traditional 
ground-based capabilities.  This contribution reviews the current status of 
optical spectroscopic observations of LINERs.  Using a recently completed 
survey of nearby galaxies, I will summarize the demographics of LINERs in 
the context of other emission-line nuclei.  Next, a number of statistical 
properties of LINERs will be compared with those of low-luminosity Seyfert 
nuclei, with the hope that such a comparison may shed some light on the 
parameters of the host galaxy nuclei that influence the various manifestations 
of nuclear activity.  Many of the properties of these faint nuclei are being 
scrutinized for the first time, and some interesting trends, not obvious or 
otherwise overlooked in older surveys, will be noted.  Finally, I address the 
fraction of LINERs that show evidence of broad-line emission akin to that seen 
in Seyfert 1 nuclei and QSOs; this subset of LINERs provides strong support 
for the hypothesis that a large fraction of all LINERs truly share the same 
physical origin as other classes of active galactic nuclei (AGNs).

Excellent reviews covering some aspects of the topics presented here, but 
based on older work, have been given by Keel (1985), Heckman (1987), and 
Filippenko (1989, 1993).  It should be emphasized at the outset that the 
material presented here refers exclusively to {\it compact} LINERs ($r$ \lax 
200 pc) found in the central region of galaxies, and not to LINER-like 
nebulosities sometimes observed in other environments (see Filippenko 
in these proceedings for an overview of these systems).

\section{Definitions and Classification of Emission-Line Nuclei}

It is important to remind ourselves of the definition of LINERs.  Although 
rigorous boundaries have little physical meaning and are, to some extent, 
arbitrary, classification is operationally necessary.  Heckman (1980b) 
originally defined LINERs strictly using the optical forbidden lines of 
oxygen: [O~II] \lamb 3727 $>$ [O~III] \lamb 5007 and [O~I] \lamb 6300 $>$ 0.33 
[O~III] \lamb 5007.  Compared with the spectra of Seyfert nuclei or H~II 
regions, the low-ionization states of oxygen in the spectra of LINERs are 
unusually strong relative to its high-ionization states.  Recognizing the 
arbitrariness of this definition, Heckman drew attention to a group of 
``transition objects'' whose spectra were intermediate between those of 
``pure'' LINERs (as defined above) and classical Seyfert nuclei.

As a consequence of the near coincidence between the ionization potentials of 
hydrogen and neutral oxygen, the collisionally-excited [O~I] line in an 
ionization-bounded nebula arises predominantly from the ``partially-ionized 
zone,'' wherein both neutral oxygen and free electrons coexist.  In addition 
to O$^0$, the conditions of the partially-ionized zone are also favorable for 
S$^+$ and N$^+$, whose ionization potentials are 23.3 eV and 29.6 eV, 
respectively.  Hence, in the absence of abundance anomalies, [N~II] \lamb\lamb 
6548, 6583 and [S~II] \lamb\lamb 6716, 6731 are strong (relative to, say, 
H\al) whenever [O~I] \lamb\lamb 6300, 6363 are strong, and vice versa.  This 
theoretical expectation and the empirical evidence that extragalactic H~II 
regions rarely exhibit [N~II] \lamb 6583 /H\al\ \gax 0.6 (e.g., Searle 
1971) have led some subsequent investigators to short-cut Heckman's 
original definition of LINERs.  For instance, it has become customary to  
classify emission-line objects solely on the basis of the [N~II]/H\al\ ratio 
(e.g., Keel 1983b; Keel \etal 1985; Phillips \etal 1986; V\'eron-Cetty \& 
V\'eron 1986).  While this convention 
does permit a convenient first-order separation between nuclei photoionized by 
stars (small [N~II]/H\al) and those photoionized by a harder, AGN-like 
spectrum (large [N~II]/H\al), it provides no information on the excitation 
level of the AGN-like objects --- in other words, one cannot distinguish 
LINERs from Seyfert nuclei.  There are two additional complications. A 
classification scheme that relies on [N~II]/H\al\ alone obviously is sensitive 
to variations in the abundance of N, which appears to be enhanced in some 
galactic nuclei (Storchi-Bergmann \& Pastoriza 1989, 1990; Ho, Filippenko, \& 
Sargent 1996d).  The net effect would be to falsely designate star-forming 
nuclei having enhanced N abundance as AGNs.  Moreover, the reliability of the 
[N~II]/H\al\ ratio depends critically on the accuracy of the separation 
between the emission and absorption components of the H\al\ line.  Although 
the ability to model and remove the stellar contribution to the integrated 
spectra is an inherent limitation to any method of classification (see \S\ 3 
and \S\ 4), it is preferable to use as many line ratios as possible to 
strengthen confidence in the classification assignment.

In the work to be discussed below, I will be using the classification 
criteria advocated by Veilleux \& Osterbrock (1987), which are motivated in 
part by the principles of Baldwin, Phillips, \& Terlevich (1981).  Based on 
the dereddened line-intensity ratios [O~III] \lamb 5007/H\bet, 
[O~I] \lamb 6300/H\al, [N~II] \lamb 6583/H\al, and [S~II] \lamb\lamb 6716, 
6731/H\al\ (H\bet\ and H\al\ refer only to the narrow component of the line), 
the Veilleux-Osterbrock system is not only relatively insensitive to 
extinction corrections, but also conveniently falls within the spectral 
range of the optical survey to be described in \S\ 4.  
For concreteness, the following definitions will be adopted:
H~II nuclei ([O~I] $<$ 0.08 H\al, [N~II] $<$ 0.6 H\al, [S~II] $<$ 0.4 H\al),
Seyferts ([O~I] $\geq$ 0.08 H\al, [N~II] $\geq$ 0.6 H\al, [S~II] $\geq$ 0.4
H\al, [O~III]/H\bet\ $\geq$ 3), and LINERs ([O~I] $\geq$ 0.17 H\al, [N~II]
$\geq$ 0.6 H\al, [S~II] $\geq$ 0.4 H\al, [O~III]/H\bet\ $<$ 3).  Although the 
adopted definition of LINERs differs from that of Heckman, inspection of the 
full optical spectra of Ho, Filippenko, \& Sargent (1993) reveals
that emission-line nuclei classified as LINERs based on the Veilleux \&
Osterbrock diagrams almost invariably also satisfy Heckman's criteria.  This
is a consequence of the inverse correlation between [O~III]/H\bet\ and
[O~II]/[O~III] in photoionized gas with fairly low excitation ([O~III]/H\bet\
\lax 3; see Fig. 2 in Baldwin \etal 1981).
 
In addition to these three categories of nuclei, Ho \etal (1993) identified a
class of ``transition objects'' (in retrospect, a poor choice of terminology) 
whose [O~I] strengths are intermediate between those of H~II nuclei and 
LINERs.  Although O-star models with an appropriate choice of parameters can 
account for their line-intensity ratios of these objects
(Filippenko \& Terlevich 1992), an alternative explanation is that these
objects are composite systems having both an H~II region and a LINER component
(Ho \etal 1993).  We will define transition objects using the same criteria as
for LINERs, except that 0.08 H\al\ $\leq$ [O~I] $<$ 0.17 H\al.  
 
\begin{figure}
\psfig{file=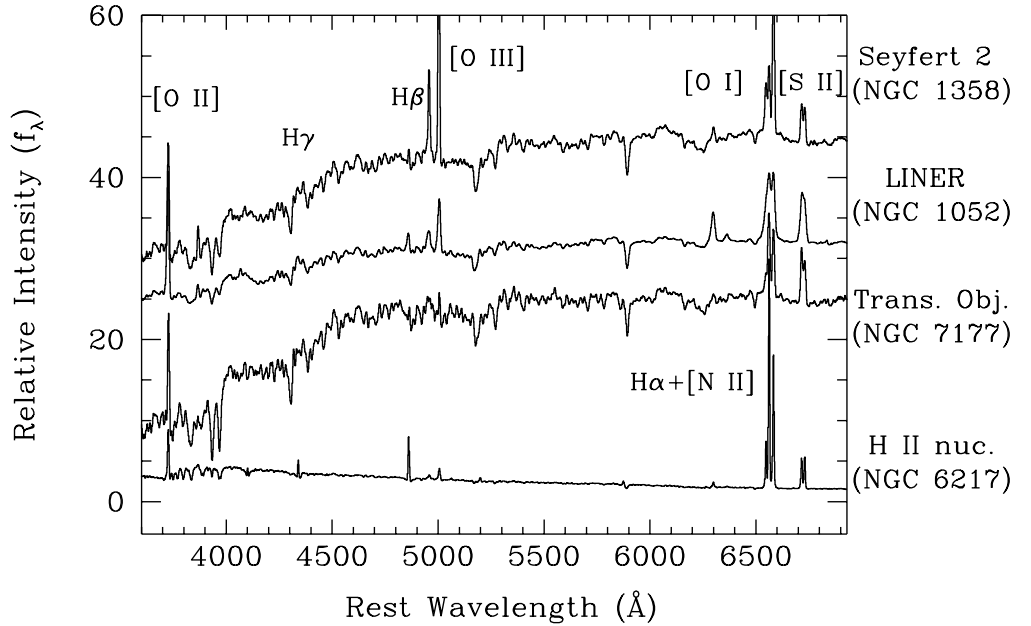,width=5.5truein,angle=-90}
\caption{Sample optical spectra of the various classes of emission-line 
nuclei.} \label{fig1}
\end{figure}

It should be emphasized that the classification process is not always
straightforward, since the three conditions involving the low-ionization
lines do not hold simultaneously in all cases.  In view of potential selective
N enhancement in galactic nuclei, less weight is given to the [N~II]/H\al\ 
ratio than to either [O~I]/H\al\ or [S~II]/H\al.  [O~I]/H\al, if reliably 
determined, deserves the most weight, since it is most sensitive to the shape 
of the ionizing spectrum.  Figure~\ref{fig1}
shows sample spectra of the various classes of objects outlined above.

\section{Previous Surveys}
It was apparent from some of the earliest redshift surveys that the 
central regions of galaxies often show evidence of strong emission lines 
(e.g., Humason, Mayall, \& Sandage 1956).  A number of studies also indicated 
that in many instances the spectra revealed abnormal line-intensity ratios, 
most notably the unusually great strength of [N~II] relative to H\al\ 
(Burbidge \& Burbidge 1962, 1965; Rubin \& Ford 1971).  That the optical 
emission-line spectra of some nuclei show patterns of low ionization was 
recognized from time to time, primarily by Osterbrock and 
his colleagues (e.g., Osterbrock \& Dufour 1973; Osterbrock \& Miller 1975; 
Koski \& Osterbrock 1976; Costero \& Osterbrock 1977; Grandi \& Osterbrock 
1978; Phillips 1979), but also by others (e.g., Disney \& Cromwell 1971; 
Fosbury \etal 1977, 1978; Danziger, Fosbury, \& Penston 1977; Penston \& 
Fosbury 1978; Stauffer \& Spinrad 1979).

Most of the activity in this field culminated in the 1980s, beginning with the 
recognition (Heckman, Balick, \& Crane 1980; Heckman 1980b) of LINERs 
as a major constituent of the extragalactic population, and followed by further 
systematic studies of larger samples of galaxies (Stauffer 1982a, b; Keel 
1983a, b; Filippenko \& Sargent 1985; Phillips \etal 1986; V\'eron \& 
V\'eron-Cetty 1986; V\'eron-Cetty \& V\'eron 1986; see summary in Table 1).  
At optical wavelengths the nuclear component in a ``normal'' galaxy is 
generally much weaker than the stellar background of its bulge.  In addition 
to having very small equivalent widths, many of the emission lines are 
severely blended and diluted by stellar absorption lines.  Thus, adequate 
removal of the stellar contribution to the integrated spectrum is an absolute 
prerequisite to any quantitative analysis of the emission-line component for 
the sources in question.  In this regard, with the exception of the survey by 
Filippenko \& Sargent (which will be considered later), the rest suffer from 
several major drawbacks (Table 1).  Although most of the surveys attempted 
some form of starlight subtraction, the accuracy of the methods used tended to 
be fairly limited (see discussion in Ho, Filippenko, \& Sargent 1996c), the 
procedure was sometimes inconsistently applied, and in 
two of the surveys subtraction was largely neglected.  The problem is 
exacerbated by the fact that the effective aperture used for the observations 
was quite large, thereby admitting an unnecessarily large amount of 
starlight.  Furthermore, most of the data were collected with rather poor 
spectral resolution ($\Delta \lambda\,\approx$ 10 \AA).  Besides losing useful 
kinematic information, severe blending between the emission and absorption 
components further compromises the ability to separate the two.

\begin{table}
\caption{Optical Spectroscopic Surveys\tablenotemark{a} \ of Nearby Galaxy Nuclei}\label{tbl-1}
\begin{center}\scriptsize
\begin{tabular}{crllcccrl}
\tableline
& No. & "$\delta$ & Types & "$B$ & Aper. & \lamb\lamb& 
$\Delta \lambda$ & Starlight \\
& & (deg) & & (mag) & "($^{\prime\prime}$) & (\AA) & (\AA) & 
Subtraction\\
\tableline

H &    93 & $\geq$+40           &  All      & 
12.0     &  6 & 3500--5300\tablenotemark{b} &  8 & Template           \\
S &   139 & $\geq$--30,$\leq$+60& Spirals   & 
13.0     &  8 & 4700--7200     & 10 & None\tablenotemark{c}               \\
K &    93 & $\geq$--15,$\leq$+40& S0/a--Scd & 
12.0     &  8 & 5400--7600     & 10 & Templ.+synthesis \\
V &   320 & $<$+20             &  E--Sc    & 
---\tablenotemark{d} &  4 & 4000--7000\tablenotemark{e}     
& 10\tablenotemark{e} & None or statistical\\
P &   203 & $\leq$--32          &  E, S0    & 
14.0     &  2 & 6100--7100     &  3 & Average template  \\
\end{tabular}
\end{center}
\tablenotetext{a}{H = Heckman et al. 1980; S = Stauffer 1982a; K = Keel 1983a; 
V = V\'eron-Cetty \& V\'eron 1986; P = Phillips \etal 1986.}
\tablenotetext{b}{A subset of 22 galaxies was observed in the spectral 
range 5100--6900 \AA\ with $\Delta \lambda$ = 6 \AA.}
\tablenotetext{c}{Template subtraction applied to a small subset.}
\tablenotetext{d}{$M_B$ $\leq$ --21 mag and $cz\,<\,3000$ \kms.}
\tablenotetext{e}{A subset of 149 galaxies was observed around H\al\ with 
$\Delta \lambda$ = 4 \AA.}
\end{table}

%Salzer, Caldwell
 
\section{The Palomar Survey of Nearby Galaxies}

Despite the sizable observational effort invested during the last decade, it 
is clear that much can be gained from a survey having greater sensitivity to 
the detection of emission lines.  The sensitivity can be improved in at 
least four ways --- by taking spectra with higher signal-to-noise ratio 
(S/N) and spectral resolution, by using a narrower slit to better isolate the 
nucleus, and by employing more effective methods to handle the starlight 
correction.

\begin{figure}
\hskip 1.5truein
\psfig{file=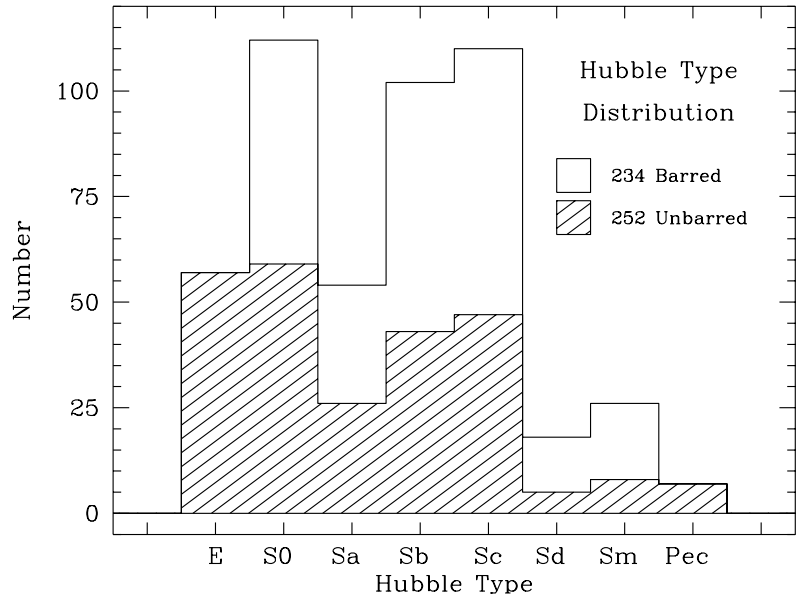,width=4.5truein,angle=-90}
\caption{Distribution of Hubble types for the 486 galaxies in the survey.
Ordinary (unbarred) galaxies are shown in the hatched histogram, and barred 
(SB and SAB) galaxies in the unhatched histogram.  Classifications taken 
from de Vaucouleurs \etal (1991).} \label{fig2}
\end{figure}

Using a double CCD spectrograph mounted on the Hale 5-m reflector at Palomar 
Observatory, high quality, moderate-resolution, long-slit spectra were 
obtained for a magnitude-limited ($B_T\,\leq$ 12.5 mag) sample of 486 northern 
($\delta$ $>$ 0\deg) galaxies (Filippenko \& Sargent 1985, 1986; Ho, 
Filippenko, \& Sargent 1995).  The red camera covered the range 
6210--6860~\AA\ with $\sim$2.5~\AA\ resolution, while the corresponding values 
for the blue camera were 4230--5110~\AA\ and $\sim$4~\AA.  Most of the 
observations were obtained with a narrow slit (generally 2\asec, and 
occasionally 1\asec), and the exposure times were suitably long (up to 1 hr or 
more for some objects with low central surface brightness) to secure data of 
high S/N.  This survey contains the largest data base of homogeneous and 
high-quality optical spectra of nearby galaxies yet published.  The selection 
criteria of the survey ensure that the sample is a good representation of the 
local ($z\,\approx\,0$) galaxy population (Fig.~\ref{fig2}), 
and the proximity of the objects (median distance = 17 Mpc) enables fairly good 
spatial resolution to be achieved (typically \lax 200 pc).

A common strategy for removing the starlight from an integrated spectrum 
is that of ``template subtraction,'' whereby a template spectrum devoid of
emission lines is suitably scaled to, and subtracted from, the spectrum of
interest to yield a continuum-subtracted, pure emission-line spectrum (e.g.,
Costero \& Osterbrock 1977; Filippenko \& Halpern 1984; Filippenko \& Sargent 
1988; Ho \etal 1993).  In practice, the template is derived either from the 
spectrum of a different galaxy or from the spectrum of an off-nuclear position 
in the same galaxy.  This approach, however, suffers from some limitations.  
For instance, the absorption-line galaxy chosen as the template may not 
exactly match the stellar component of the object in question; previous 
studies generally invested limited observing time to the acquisition of
template spectra.  In the case where an off-nuclear spectrum is used as the 
model, it may not be completely free of emission, and one cannot be sure that 
radial gradients in the stellar population are absent.

\begin{figure}
\psfig{file=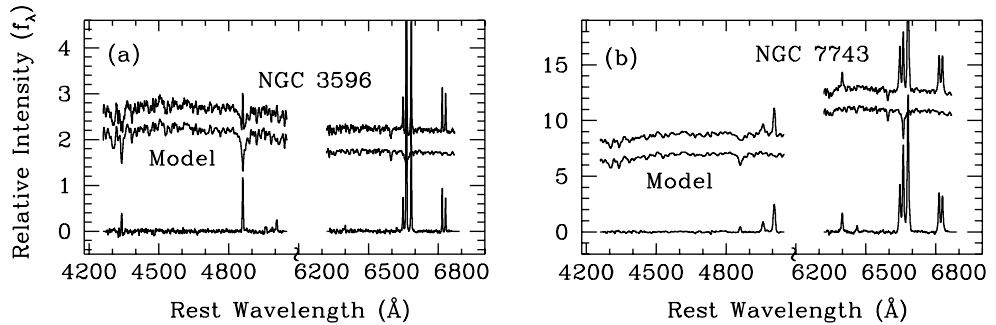,width=5.5truein,angle=-90}
\caption{Illustration of the method of starlight subtraction.  In each
panel, the top plot shows the observed spectrum, the middle plot the
best-fitting ``template'' used to match the stellar component, and the
bottom plot the difference between the object spectrum and the template.
In the case of NGC~3596 ({\it a}), the model was constructed from NGC~205 and
NGC~4339, while for NGC~7743 ({\it b}), the model was derived from a linear
combination of NGC~205, NGC~4339, and NGC~628.} \label{fig3}
\end{figure}

To perform the starlight subtraction in a more objective and efficient manner 
than has been done in the past, a modified version of the template-subtraction 
technique that takes advantage of the large number of absorption-line galaxies 
in the survey was developed (Ho \etal 1996c).  Given a list 
of input template spectra and an initial guess of the velocity dispersion, the 
$\chi^2$-minimization algorithm of Rix \& White (1992) solves for the systemic 
velocity, the line-broadening velocity dispersion, the relative contributions 
of the various templates, and the general continuum shape.  The best-fitting 
model is then subtracted from the original spectrum, yielding a pure 
emission-line spectrum.  Figure~\ref{fig3}
illustrates this process for the H~II nucleus in NGC~3596 and for the 
Seyfert 2 nucleus in NGC~7743.  In the case of NGC~3596, the model consisted 
of the combination of the spectrum of NGC~205, a dE5 galaxy with a substantial 
population of A stars, and NGC~4339, an E0 having a K-giant spectrum.  Note 
that in the original observed spectrum (top), H$\gamma$, [O~III] \lamb\lamb 
4959, 5007, and [O~I] \lamb 6300 were hardly visible, whereas after starlight 
subtraction (bottom) they can be easily measured.  The intensities of both 
H\bet\ and H\al\ have been modified substantially, and the ratio of the two 
[S~II] \lamb\lamb 6716, 6731 lines changed.  The effective template for 
NGC~7743 made use of NGC~205, NGC~4339, and NGC~628, an Sc galaxy with a 
nucleus dominated by A and F stars.

\section{Demographics of Emission-Line Galaxies}

Although the specific numbers cited differ from one investigator to another, 
all the older surveys discussed in \S\ 3 agree that LINERs are extremely 
common in nearby galaxies.  They also concur that the detection rate of LINERs 
varies strongly with Hubble type, with early-type systems 
being the preferred hosts; this result essentially confirms what was already 
found by Burbidge \& Burbidge (1962), who noted that most of the galaxies 
showing enhanced [N~II]/H\al\ ratios tended to be of early type.

\begin{figure}
\psfig{file=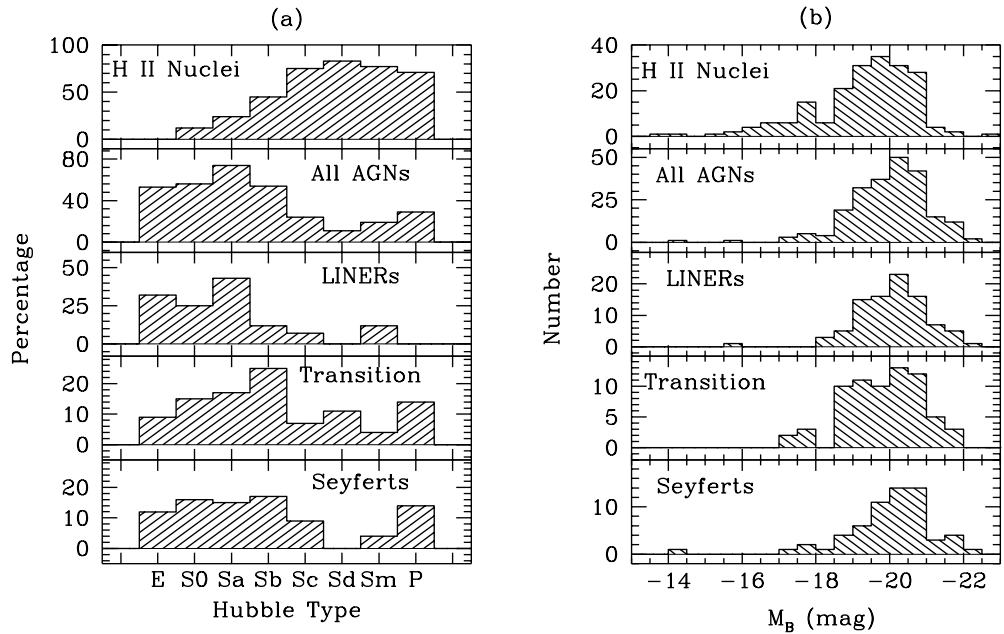,width=5.5truein,angle=-90}
\caption{({\it a}) Percentage of galaxies with the various classes of
emission-line nuclei detected as a function of Hubble type. ({\it b})
Distribution of the classes of emission-line nuclei as a function of the
absolute $B$ magnitude of the host galaxy.} \label{fig4}
\end{figure}

Not surprisingly, the Palomar survey likewise finds the same trends.  The 
important distinction, however, is that the results from the present survey 
are {\it quantitatively} much more reliable, for reasons already discussed 
in \S\ 4, both in a statistical sense as well as on an object-by-object 
basis.  The detection rates of the various classes of emission-line nuclei 
defined in \S\ 2 are given in Table~\ref{tbl2}
%
%The following are from /home/lho/projects/liners/stsci_review/activity.stats
%Updated as per page 34a of notebook B16
%
\begin{table}
\caption{Percentages of Emission-Line Nuclei in the Palomar 
Survey\tablenotemark{a},\tablenotemark{b}}\label{tbl2}
\begin{center}\scriptsize
\begin{tabular}{lccccccc}
\tableline
Hubble Type &  S & L & T & L+T & H & A & AGN \\
\tableline
All types   & 13  (62)                  & 19 \  (92)&  14  (69)&33   (161)& 40   (195)& 14  (66)& 46   (223) \\
E--E/S0     & 12 \ (7)                  & 32 \  (18)& \ 9 \ (5)&41 \  (23)&\ 0 \ \ (0)& 46  (26)& 53 \  (30) \\
S0--S0/a    & 16  (18)                  & 25 \  (28)&  15  (17)&40 \  (45)& 12 \  (13)& 32  (36)& 56 \  (63) \\
Sa--Sab     & 15 \ (8)                  & 43 \  (23)&  17 \ (9)&60 \  (32)& 24 \  (13)&\ 2 \ (1)& 74 \  (40) \\
Sb--Sbc     & 17  (17)                  & 12 \  (12)&  25  (26)&37 \  (38)& 45 \  (46)&\ 1 \ (1)& 54 \  (55) \\
Sc          &\ 9  (10)                  &\ 7 \ \ (8)& \ 7 \ (8)&14 \  (16)& 75 \  (83)&\ 1 \ (1)& 24 \  (26) \\
Scd--Sd     &\ 0 \ (0)                  &\ 0 \ \ (0)&  11 \ (2)&11 \ \ (2)& 83 \  (15)&\ 6 \ (1)& 11 \ \ (2) \\
Sdm,Sm,Im,I0&\ 4 \ (1)                  & 12 \ \ (3)& \ 4 \ (1)&16 \ \ (4)& 77 \  (20)&\ 0 \ (0)& 19 \ \ (5) \\
``S'', S pec& 14 \ (1)                  &\ 0 \ \ (0)&  14 \ (1)&14 \ \ (1)& 71 \ \ (5)&\ 0 \ (0)& 29 \ \ (2) \\
\end{tabular}
\end{center}
\tablenotetext{a}{S = Seyfert; L = LINER; T = transition object; H = H~II 
nucleus; and A = absorption-line nucleus (i.e., contains no emission lines).  
The column denoted by ``L+T'' is the sum of ``L'' and ``T,'' and ``AGN'' is 
the sum of ``S,'' ``L,'' and ``T''.  The percentages of objects having 
emission lines is just 100$-$A.}
\tablenotetext{b}{In each case, the first entry is the percentage of the total 
of that type, and the value in parentheses is the actual number of objects.}
\end{table}
and graphically illustrated in Figure~\ref{fig4}{\it a}.
The conclusions that can be drawn are the following.  
\begin{enumerate}
\item  At the limit of our survey, which is at least 4 times more 
sensitive to the detection of emission lines than any of the older surveys, 
most galaxies (86\%) exhibit optical line emission in their central few 
hundred parsecs, implying that ionized gas is almost invariably present.  This 
fraction, of course, represents a lower limit.  Keel (1983a) detected emission 
in all the galaxies he surveyed, but his sample was restricted to spirals; 
Table 2 confirms that essentially all spirals have nuclear emission lines.  
The Hubble type distribution of the surveys of Heckman \etal (1980) and 
V\'eron-Cetty \& V\'eron (1986) more closely matches that of the present 
sample, and, in these, the detection rate was only $\sim$60\%--65\%.  
\item  Seyfert nuclei can be found in at least 10\% of all galaxies with 
$B_T\,\leq$ 12.5 mag, the vast majority of which ($\sim$80\%) have early 
Hubble types (E--Sbc).  The fraction of galaxies hosting Seyfert nuclei has 
roughly doubled compared to previous estimates (Stauffer 1982b; Keel 1983b; 
Phillips, Charles, \& Baldwin 1983; Maiolino \& Rieke 1995).  It is 
interesting to note that Seyfert nuclei, at least with luminosities 
as low as those here, do {\it not} exclusively reside in spirals, as is 
usually believed (e.g., Adams 1977; Weedman 1977).   In fact, galaxies of 
types E and E/S0 have roughly the same probability of hosting a Seyfert 
nucleus as those of types between S0 and Sbc.  
\item  ``Pure'' LINERs are present in $\sim$20\% of all galaxies, whereas 
transition objects, which by assumption also contain a LINER component, account 
for another $\sim$15\%. Thus, if all LINERs can be regarded as genuine AGNs, 
they truly are the most populous constituents --- they make up $>$70\% 
of the AGN population (here taken to mean all objects classified as Seyferts, 
LINERs, and transition objects) and a full 1/3 of all galaxies.  The latter 
statistic broadly supports earlier findings by Heckman (1980b) and others.  
\item  The Hubble type distribution of ``pure'' LINERs is virtually identical 
to that of Seyferts; the same can be said for the distribution of absolute 
magnitudes (Fig. 4{\it b}), both groups having a median $M_B$ = --20.2 mag.  
On the other hand, the hosts of many transition objects apparently have 
somewhat later Hubble types and fainter absolute magnitudes (median $M_B$ = 
--20.0 mag), consistent with the idea that these systems are composites of 
``pure'' LINERs and H~II nuclei.  
\item  H~II nuclei, in striking contrast to AGNs, occur preferentially in 
late-type galaxies (Heckman 1980a; Keel 1983a; Terlevich, Melnick, \& Moles 
1987).  Quite surprisingly, not a single elliptical galaxy falls into 
this category.  This is consistent with the survey of early-type (E and 
S0) galaxies of Phillips \etal (1986); the few objects they identified 
as having H~II nuclei are all classified S0 (two are E-S0).  Narrow-band 
imaging surveys of elliptical galaxies (e.g., Shields 1991) often
reveal detectable amounts of warm (T $\approx$ 10$^4$ K) ionized gas in 
their centers.  Although the dominant ionizing agent responsible for the 
line emission is still controversial (Binette \etal 1994, and references 
therein), our failure to detect spectra resembling ordinary metal-rich 
H~II regions among the $\sim$60 ellipticals in our survey suggests that 
young massive stars are probably not the culprit, unless the physical 
conditions in the centers of ellipticals conspire to make H~II regions 
look very different from those seen in the nuclei of S0s and early-type 
spirals.
\end{enumerate}

Theoretical studies (e.g., Heller \& Shlosman 1994) suggest that large-scale 
stellar bars can be highly effective in delivering gas to the central few 
hundred parsecs of a spiral galaxy, which may then lead to rapid star 
formation.  Further instabilities may result in additional inflow to smaller 
physical scales relevant for AGNs.  Thus, provided that a reservoir of gas 
exists, the presence of a bar might be expected to influence the fueling rate, 
and hence the activity level.  Being sufficiently large and unbiased with 
respect to bar type, the Palomar survey can be used to examine this issue.  
Ho, Filippenko, \& Sargent (1996a, e) find that the presence of a bar does 
indeed enhance both the probability and rate of the formation of massive stars 
in galaxy nuclei, but only for spirals with types earlier than Sbc.  By 
contrast, AGNs seem to be altogether unaffected.

%Dynamical interactions with neighboring companions similarly should lead to 
%angular momentum transfer (e.g., Hernquist 1989).  Unfortunately, this effect 
%cannot be examined with these data, because the sample is such that most of 
%the galaxies are relatively isolated.

\section{Statistical Properties of LINERs and Seyfert Nuclei}

\subsection{Emission-Line Luminosity}

\begin{figure}
\psfig{file=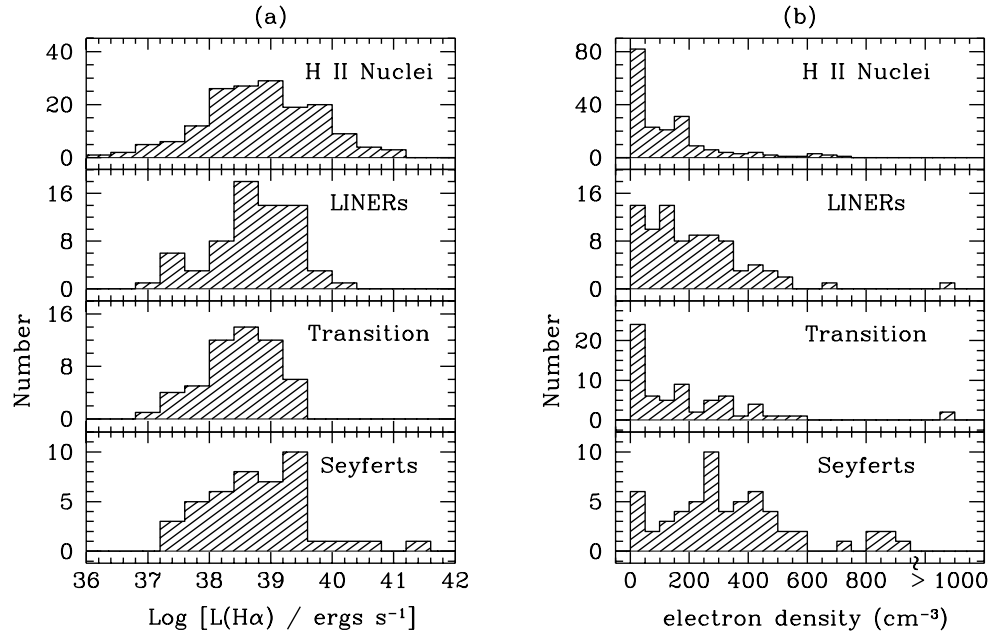,width=5.5truein,angle=-90}
\caption{({\it a}) Distribution of dereddened luminosities for the narrow 
H\al\ emission line.  ({\it b}) Distribution of electron densities derived 
from [S~II] \lamb\lamb 6716, 6731.} \label{fig5}
\end{figure}

In lieu of direct measurement of the nonstellar featureless continuum at 
optical wavelengths, an almost impossible feat for the low-luminosity sources 
in question, one might use, as a substitute, an indirect measure such as the 
luminosity of a narrow emission line powered by the continuum.  In luminous 
AGNs, whose nonstellar optical continuum generally overwhelms the stellar 
background, the H\al\ luminosity scales linearly with the luminosity of the 
continuum (Searle \& Sargent 1968; Yee 1980; Shuder 1981).  However, using 
line luminosities derived 
from slit spectroscopy is not without complications, especially for the 
relatively narrow slit employed in the Palomar survey.  For any given object, 
the amount of line emission sampled will depend on its distance as well as on
the physical extent of the line-emitting region.  Moreover, circumnuclear H~II 
regions undoubtedly contaminate the line emission at some level (indeed, for 
transition objects, this is assumed {\it a priori}).  These limitations 
notwithstanding, it can be argued that a {\it statistical} examination of such 
line luminosities might still be of value, as the individual ``fluctuations'' 
will tend to average out for sufficiently large samples.  The H\al\ 
luminosities for the various classes of emission-line nuclei are shown in 
Figure~\ref{fig5}{\it a}.
The incredible feebleness of the low-luminosity AGNs can be readily 
appreciated by realizing that a sizable fraction of H~II nuclei and disk giant 
H~II regions (e.g., Kennicutt 1988), not to mention starburst nuclei (e.g., 
Balzano 1983), in fact have much stronger H\al\ luminosities than these 
sources.  Remarkably, the distributions for LINERs and Seyferts appear very 
similar, both having a median $L$(H\al) $\approx$ 6\e{38} \lum; transition 
objects tend to be somewhat less luminous, but the difference is insignificant 
according to the Kolmogorov-Smirnov (K-S) test.  (The probability that the 
LINERs and transition objects are drawn from the same population, $P_{\rm KS}$, 
is 0.16.)  The above comparison is not obviously affected by known systematic 
biases, since all three subclasses have virtually identical distance 
distributions, modest reddening corrections were consistently applied, and 
it was already shown (\S\ 5) that the host galaxies of LINERs and Seyferts are 
grossly similar.  

Heckman (1980b) claimed that the line luminosities of LINERs are smaller 
than those of Seyferts, contrary to what we and Stauffer (1982b) find.  The 
principal reason for this difference appears to be that Heckman used for 
comparison a sample of Seyferts, taken from Adams \& Weedman (1975), which 
was biased toward luminous sources.

One should be cautious in interpreting these results.  Superficially, it 
appears that both LINERs and Seyferts have the same ``level'' of activity, 
if one assumes that the narrow-line luminosity is an appropriate yard stick 
for gauging the power output of the central source.   It remains to be 
demonstrated, however, that the line-continuum relation of luminous AGNs 
continues to hold for low-luminosity sources.  There is preliminary evidence, 
for instance, that the spectral energy distributions of LINERs and 
low-luminosity Seyferts may differ appreciably from those of high-luminosity 
sources (Ho, Filippenko, \& Sargent 1996b).  If so, the equivalent widths of 
emission lines, and hence the slope of the line-continuum relation, will 
vary systematically with luminosity.

\subsection{Electron Density}

The electron density can be estimated from the ratio of the two [S~II] lines, 
at least for the portion of the narrow-line region (NLR) characterized by 
densities not greatly in excess of the critical density of [S~II] 
($\sim$3\e{3} \cc), above which the lines become collisionally de-excited.  
As shown in \S\ 6.4, a range of densities, spanning nearly five orders of 
magnitude, exists in the NLRs of some LINERs and Seyferts.  The [S~II] 
densitometer strictly probes only the low-density regions.  Figure 5{\it b} 
indicates that LINERs have {\it smaller} electron densities (median $n_e$ = 
175 \cc) than Seyferts (median $n_e$ = 290 \cc), and the difference is highly 
significant according to the K-S test ($P_{\rm KS}$ = 0.00078).  Transition 
objects have smaller densities than LINERs, most notably in a considerable 
excess of low-density members, as seen in a large fraction of H~II nuclei.  

These density measurements are in substantial disagreement with previous 
studies, which typically find densities on the order of 1000 \cc\ (Stauffer 
1982b; Keel 1983b; Phillips \etal 1986).  The discrepancy can be traced 
to a common culprit, which serves as an excellent lesson in the pitfalls of 
measuring weak lines in galaxy nuclei.  Careful inspection of Figure 3 
reveals that [S~II] \lamb 6716 is affected by a stellar absorption line
due to Ca~I \lamb 6718 [see also Fig. 12 of Filippenko \& Sargent (1985)].  
Since the emission lines in the sources of interest generally have very low 
equivalent widths (typically 2--3 \AA), the absorption feature, though weak, 
can significantly depress the ratio of [S~II] \lamb 6716 to [S~II] \lamb 6731; 
this effect will artificially raise the derived electron density.  The older 
studies in question almost never performed starlight subtraction to the degree 
of precision required to notice this, thereby systematically overestimating 
the inferred densities.

It is interesting to point out that the electron densities among Seyfert 
nuclei appear to decrease with decreasing nuclear luminosity.  In a sample 
of bright, mostly Markarian Seyfert 2 galaxies, Koski (1978) found that the 
average density, again as determined from [S~II], is $\sim$2000 \cc, far 
greater than that encountered in the present sample of low-luminosity 
Seyferts.  Although the systematic effect discussed above may also affect 
Koski's measurements to some degree, it probably cannot account for the 
large difference, especially in view of the much larger emission-line 
equivalent widths in his sample.  The same trend of density variation is also 
clearly seen in the smaller sample of low-luminosity sources published by 
Phillips \etal (1983); for the 17 objects in which both [S~II] lines were 
tabulated, I calculate $\langle n_e \rangle$ = 340 \cc.

\subsection{Internal Reddening and Inclination Effects}

Another parameter that can be easily examined is the internal reddening along 
the line of sight, as inferred from the relative intensities of the narrow 
Balmer emission lines.  The conventional Balmer decrement method, 
unfortunately, assumes that the extinction arises from a uniform, foreground 
screen of dust, and it is unclear to what extent such an oversimplified 
geometry applies to the actual line-emitting regions in galaxy nuclei.  The 
derived reddening values, therefore, should be strictly regarded as lower 
limits.  With this caveat in mind, it is intriguing that LINERs are noticeably 
{\it less} reddened than Seyferts (Fig. 6{\it a}; 
$P_{\rm KS}$ = 0.0023).  That LINERs are also less reddened compared to 
transition objects is to be expected, since H~II nuclei in general are much 
more heavily extinguished than LINERs [median $E(B-V)$ = 0.21 and 0.47 mag for 
LINERs and H~II nuclei, respectively].  These data constitute the first set of 
reliable reddening measurements for such faint nuclei.  In the older surveys, 
the Balmer decrements were either completely unconstrained (e.g., because only 
the red part of the spectrum was surveyed) or otherwise very poorly determined 
because of the difficulties associated with starlight correction.  Heckman's 
(1980b) suspicion that the H\al/H\bet\ ratios in LINERs may be intrinsically 
and abnormally high is not supported by the present observations.

\begin{figure}   
\psfig{file=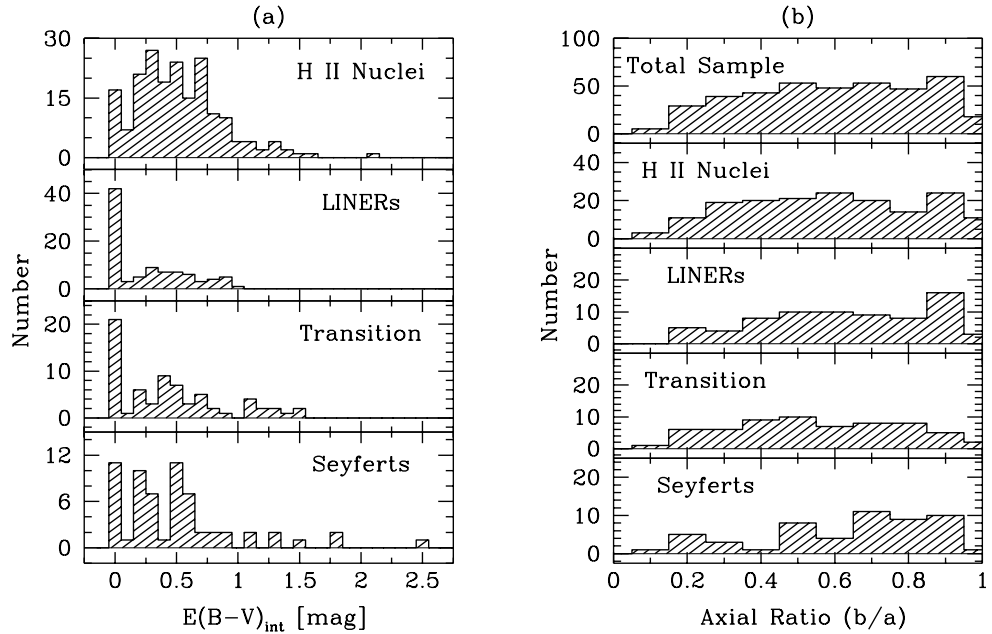,width=5.5truein,angle=-90}
\caption{({\it a}) Distribution of internal reddening as measured from
H\al/H\bet.  ({\it b}) Distribution of axial ratios (b/a = minor-axis
diameter/major-axis diameter) taken from de Vaucouleurs \etal (1991); galaxies
classified as E, Sdm, Sm, Im, as well as those with highly uncertain
classifications, have been excluded.}\label{fig6}
\end{figure}

How are these results to be interpreted?  It is highly instructive to consider
the axial ratios, or inclination angles, of the host galaxies for the various 
subclasses (Fig. 6{\it b}).  Binney \& de Vaucouleurs (1981) find that the 
distribution of apparent axial ratios ($b/a$) of S0s and spirals (Sa--Sd) in 
the RC2 is 
essentially flat between 0.2 and 1.  The axial ratios for the entire Palomar 
sample, after excluding ellipticals, Magellanic irregulars (Sdm, Sm, Im), and 
objects with highly distorted morphologies, are shown in the upper panel of 
Figure~\ref{fig6}{\it b}.  
Note that the distribution has a slight 
positive slope; this property is expected in a magnitude-limited sample, since 
internal absorption tends to shift edge-on systems above the limiting magnitude
(Maiolino \& Rieke 1995).  The distributions of axial ratios among the 
various subclasses are quite similar, but some subtle differences can be 
discerned.  Although Seyferts are found in hosts of all inclinations, there 
appears to be a preference for systems viewed more face-on (larger $b/a$).  
Such an inclination bias is well known in other samples of Seyfert galaxies 
(Keel 1980; McLeod \& Rieke 1995; Maiolino \& Rieke 1995), but it is much less 
pronounced in the present sample because of our ability to detect weak 
emission lines.   There is also some indication that a similar selection 
effect is present in LINERs, and, at any rate, Seyferts and LINERs do not have 
different inclinations.  Thus, from geometric considerations alone, these 
two groups should be equally reddened.  The enhanced reddening observed in 
Seyferts relative to LINERs, therefore, must point to subtle, intrinsic 
differences present in the physical conditions of their NLRs (\S\ 9).

The axial ratios of H~II nuclei and transition objects, on the other hand, 
appear to be evenly distributed, and do not seem to be affected by any obvious 
inclination bias.  It is tempting to postulate that at least some transition 
objects are nothing more than highly-inclined LINERs.  The increased path 
length along the line of sight to the nucleus increases the likelihood of 
intersecting discrete circumnuclear H~II regions as well as general extended 
nebulosity photoionized by massive stars (see also discussion in \S\ 8).  

I note, in passing, that the inclination bias among Seyfert galaxies in our 
sample is present in Seyfert 1s {\it and} Seyfert 2s, in agreement with 
McLeod \& Rieke (1995) and Maiolino \& Rieke (1995), although in our sample 
and in that of Maiolino \& Rieke the axial-ratio cutoff is not as pronounced 
as that of the CfA Seyferts studied by McLeod \& Rieke.  The larger number of 
high-inclination Seyferts detected compared to the CfA sample probably 
reflects the much larger average distance of the latter sample.  That the 
absorbing material evidently affects both the broad-line region (BLR) and the 
NLR implies that the source of obscuration lies in a flat configuration aligned 
with the plane of the galactic disk.  Based on the abrupt fall off of 
objects with $b/a$ \lax 0.5 in the CfA sample, McLeod \& Rieke argue that the 
obscuring for the existence of an ``outer torus,'' whose thickness and radial 
extent are comparable to the dimensions of the NLR ($r\,\approx$ 100 pc).  
It is unclear whether this model is applicable to the low-luminosities 
sources considered here, since opacity from the dusty disk of the galaxies 
alone can explain the inclination bias observed.

\subsection{Line Profiles and Kinematics}
 
The kinematic information contained in line profiles provides unique clues to
the LINER puzzle.  However, aside from a small handful of 
relatively crude line width measurements (e.g., Dahari \& De Robertis 1988; 
Whittle 1993, and references therein), little else is known about the line 
profiles of LINERs as a class.  Indeed, with a few exceptions, the FWHM of the 
forbidden lines in LINERs rarely exceed 500 \kms, the typical resolution of 
many of the older surveys.   Since the initial study of Heckman (1980b), 
it has commonly been assumed that the line widths of LINERs are roughly 
comparable to those of Seyferts (Wilson \& Heckman 1985; Whittle 1985, 1993), 
although Stauffer (1982b) has remarked, admittedly based on very small 
number statistics, that LINERs seem to have {\it broader} lines than Seyferts. 
The study of Phillips \etal (1986), whose spectral resolution is comparable 
to that of the Palomar survey, also has clearly shown that the typical 
line widths in LINERs are substantially smaller than what Heckman had first 
thought.

\begin{figure}
\hskip 0.75truein
\psfig{file=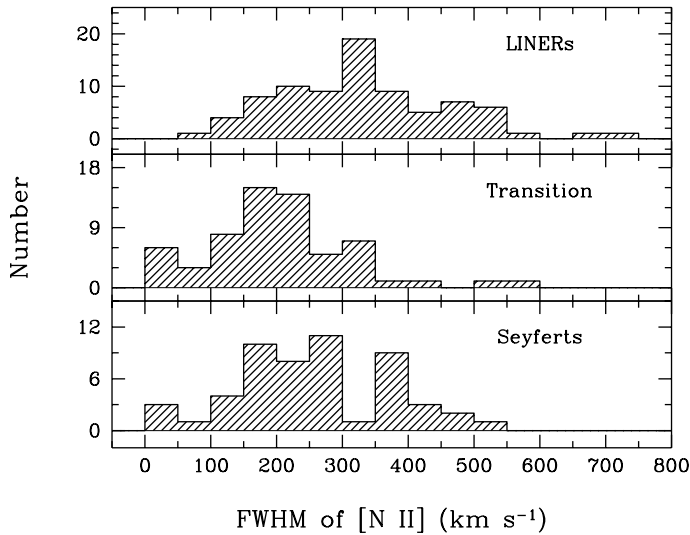,width=4.0truein,angle=0}
\caption{Distribution of FWHM of [N~II] \lamb 6583 for 88 LINERs, 62 transition
objects, and 55 Seyfert nuclei.  The line widths have been corrected for
instrumental broadening, and highly uncertain measurements have been 
excluded.} \label{fig7}
\end{figure}

Let us reexamine some of these issues using the new data set at hand.  Despite 
being blended with H\al\ most of the time, I will use [N~II] \lamb 6583 as the 
fiducial probe of the velocity field of the NLR, since it is usually the 
strongest line in the red spectrum, and it is relatively unaffected by stellar 
absorption.  [O~III] \lamb 5007 normally is more ideal for measurement of 
narrow-line profiles, but, in our case, both the S/N and the resolution of the 
blue spectra are lower than those of the red spectra.  The line widths range 
from being unresolved (\lax 115 \kms) to 500--700 \kms, with a median value 
(excluding the first bin, whose values are very uncertain because they are 
near the resolution limit) of 350, 230, and 290 \kms, respectively, for 
LINERs, transition objects, and Seyferts (Fig.~\ref{fig7}).

The characteristic line widths of LINERs reported here are much smaller
than those found by Heckman (1980b), and generally consistent with those from 
other moderate-resolution studies.  Not surprisingly, transition objects have 
narrower lines compared to LINERs; this is to be expected because of the 
difference in their average Hubble types and the well-known dependence of 
nebular line width on bulge prominence (e.g., Whittle 1992a, b).  What is 
unexpected is the clear difference evident between LINERs and Seyferts: LINERs 
have wider forbidden lines than Seyferts, significant at a level greater than 
99.999\% according to the K-S test.  

\begin{figure}
\hskip 0.75truein
\psfig{file=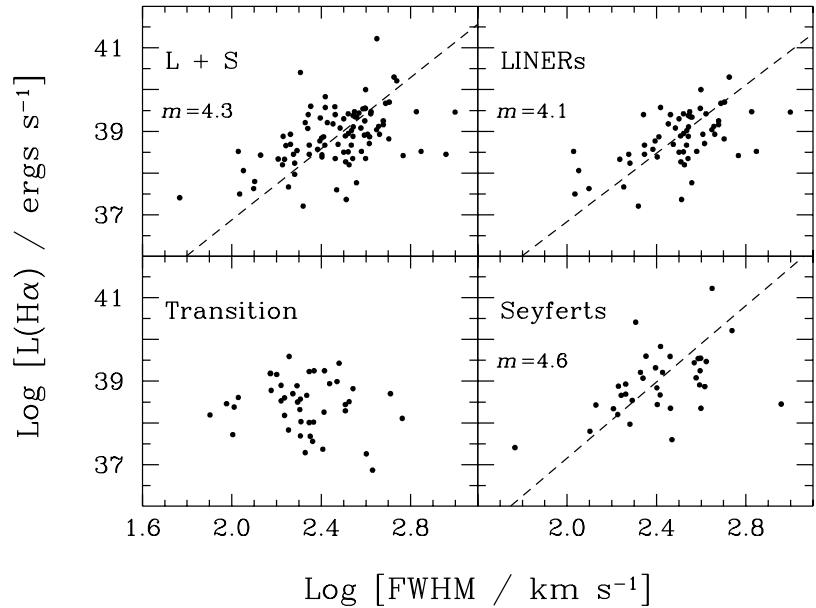,width=5.0truein,angle=-90}
\caption{Correlation between H\al\ luminosity and the FWHM of [N~II] \lamb
6583.  The line luminosities have been dereddened, and the line widths have
been corrected for instrumental resolution.} \label{fig8}
\end{figure}

Since it was first pointed out by Phillips \etal (1983), it has been 
well established that the luminosities of the forbidden lines in Seyfert nuclei
are positively correlated with their widths (Whittle 1985, 1992b).  
The Seyferts in our sample similarly obey this correlation (Fig. 8), 
although, interestingly, the slope of the correlation (4.6) is somewhat 
shallower than that of the more luminous Seyferts in Whittle's (1992b) sample 
(slope $\approx$ 5.5).  LINERs evidently also obey the correlation, contrary to what Wilson \& Heckman (1985) thought; the shallower slope (4.1) reflects the 
larger line widths found in LINERs.  Transition objects, on the other hand, 
appear not to follow the correlation, although it is likely that the 
correlation has been partially masked by our inability to completely resolve 
the narrower lines in this class of objects.  The interpretation of the 
relation between line luminosity and line width has been unclear, mainly 
because of the existence of other mutual correlations between line width, line 
luminosity, and radio power (Wilson \& Heckman 1985).  The recent analysis 
by Whittle 
(1992b), however, shows quite convincingly that the fundamental parameter 
underlying all these correlations is the bulge mass (or central gravitational 
potential) of the host galaxy.  

In light of the dependence of line width on luminosity, it is hardly 
surprising that the ``typical'' Seyfert nucleus has much narrower lines 
than conventionally assumed.  Hence, the criterion for distinguishing 
Seyfert 2 nuclei from ``normal'' emission-line nuclei (i.e., H~II nuclei) 
on the basis of the widths of the narrow lines, either as originally proposed 
by Weedman (1970, 1977), or as later modified by Balzano \& Weedman (1981) and 
Shuder \& Osterbrock (1981), is clearly inappropriate for the majority of the 
Seyfert galaxy population and should be abandoned.

Keel (1983b) found that in his sample the widths of the forbidden lines are
well correlated with galaxy inclination, implying that motion in the plane of 
galaxy disks dominates the velocity field of the NLR.  The present data set 
does not support this conclusion; no significant correlation between 
FWHM([N~II]) and galaxy axial ratio is seen for any of the subclasses of 
nuclei, or for all the subclasses added together.  Other studies have come to
the same conclusion (Heckman \etal 1981; Wilson \& Heckman 1985; Whittle 
1985; V\'eron \& V\'eron-Cetty 1986).  One can make the inference that 
either the NLR does not have a disk-like geometry in the plane of the 
galactic disk, or that the component of the velocity field in the 
galactic plane contributes only a portion of the total observed line widths 
(Whittle 1985, 1992a).

Of course, the FWHM is the crudest, first-order characterization of the line
profile.  Actually, the shapes of the emission lines in most emission-line
nuclei, when examined with sufficient spectral resolution (e.g., Heckman \etal
1981; Whittle 1985; Veilleux 1991; Ho \etal 1996f), deviate far from simple
analytic functions (such as a Gaussian), often exhibiting weak extended wings
and asymmetry.  In fact, most Seyfert nuclei have asymmetric narrow lines, and
there seems to be a preponderance of blue wings, usually interpreted as 
evidence of a substantial radial component in the velocity field coupled with 
a source of dust opacity.  It would be highly instructive to see if this trend 
extends to LINERs, as it could offer insights into possible differences 
between the NLRs in the two types of objects.  These subtleties have never 
before been examined systematically in LINERs.
 
\begin{figure}
\psfig{file=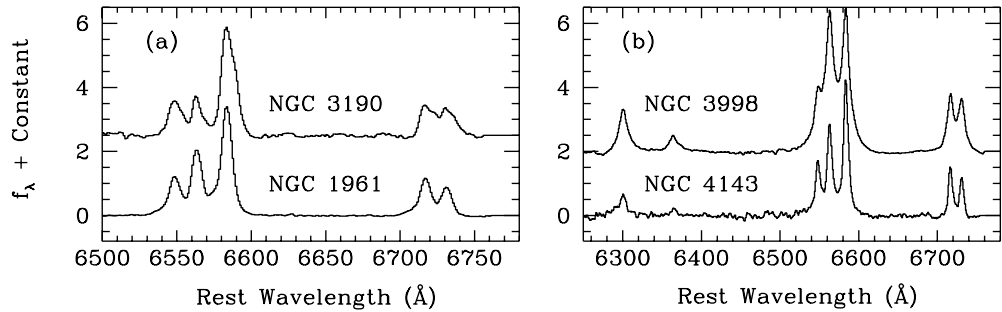,width=5.5truein,angle=-90}
\caption{({\it a}) Examples of LINERs showing line profiles with blue (NGC~1961)
and red (NGC~3190) asymmetry.  ({\it b}) Examples of LINERs having forbidden
lines whose widths vary with critical density.  Note that in both cases
[O~I] is broader than [S~II].} \label{fig9}
\end{figure}

The majority of the LINERs in our survey have emission-line spectra of
adequate S/N such that possible profile asymmetries (at, say, 20\% of the peak
intensity) can be discerned (Fig.~\ref{fig9}{\it a}).
Among these, $\sim$30\% are symmetric, $\sim$50\% have blue asymmetry, and
$\sim$20\% have red asymmetry.  Seyferts do not show any obvious differences
compared to LINERs, although a much larger proportion of transition objects
($\sim$60\%) are observed to have symmetric line profiles.  The latter finding
is probably insignificant, since it is more difficult to identify profile
asymmetries in objects having narrower lines.  From these results, one can 
conclude that (1) both LINERs and Seyferts seem to exhibit similar trends in 
their narrow-line asymmetries and that (2) when present, the sense of the 
asymmetry is preferentially to the blue.
%Another intriguing
%result is that the fraction of objects with red asymmetry, while certainly
%less than those of the opposite sense, is not negligible.
 
Detailed studies of Seyferts (e.g., De Robertis \& Osterbrock 1984, 1986) and 
LINERs (Filippenko \& Halpern 1984; Filippenko 1985; Filippenko \& Sargent 
1988; Ho \etal 1993, 1996b) in the past have found that the widths of the 
forbidden lines correlate positively with their critical densities.  This 
empirical trend has been interpreted as evidence that the NLR contains a wide 
range of gas densities (10$^2$--10$^7$ \cc), stratified such that the densest 
material is located closest to the center.  In such a picture, [O~I] \lamb 
6300 ($n_{crit}\, \approx\,10^6$ \cc) should be {\it broader} than [S~II] 
\lamb\lamb 6716, 6731 ($n_{crit}\,\approx$ 3\e{3} \cc).
 
Among the objects with securely determined FWHM for [O~I] and [S~II],
approximately 15\%--20\% of LINERs and 10\% of Seyferts show detectable evidence
of density stratification in the sense that FWHM([O~I]) $>$ FWHM([S~II]) (see 
Fig.~9{\it b} for examples).  In no instance is [O~I] ever observed to be 
narrower than [S~II].  However, these numbers need to be interpreted with 
caution.  They do {\it not} imply that objects failing to show such profile 
differences do not have density stratification, since a number of effects can 
conspire to hide this observational signature (Whittle 1985).  Furthermore, 
one's ability to discern such profile differences depends strongly on the S/N 
of the data (and on the resolution compared with FWHM), and undoubtedly many 
objects have escaped notice because of this observational selection effect.
 
Whittle (1985) finds that Seyfert 1 nuclei have a greater likelihood of
showing profile differences in their forbidden lines than do Seyfert 2s.  The
implication is that somehow density stratification in the NLR is directly
related to the presence of a BLR.  In the present sample, the same trend
seems to hold (see also Ho \etal 1993), in that, among those objects with
detectable profile differences between [O~I] and [S~II], $\sim$50\% of the
LINERs and $\sim$80\% of the Seyferts have broad H\al\ emission, significantly 
higher than the respective detection rates of broad H\al\ in the whole sample 
(\S\ 7).  But, once again, this result is difficult to evaluate, since 
selection effects heavily favor the detection of both of these traits in 
objects having data of high S/N. 

\section{Searching for Broad H\al\ Emission}

Some LINERs are known to exhibit broad H\al\ emission (FWHM of a few
thousand \kms), reminiscent of the broad emission lines that define type 1
Seyfert nuclei (Khachikian \& Weedman 1974).  This subset of LINERs suffers
from the least degree of ambiguity in physical origin and can be most safely
regarded as representing {\it genuine} low-luminosity AGNs.  These objects 
are analogous to the so-called intermediate Seyferts (types 1.8 and 1.9) 
in the terminology of Osterbrock (1981), except that their narrow-line spectra 
have low ionization and satisfy the definition of LINERs.  The luminosities
of broad H\al\ can be orders of magnitude fainter than those in
classical Seyfert 1 nuclei.  The well-known example of the nucleus of M81 
(Peimbert \& Torres-Peimbert 1981; Shuder \& Osterbrock 1981; Filippenko \& 
Sargent 1988), for example, has a broad H\al\ luminosity of only 
1.8 $\times\,10^{39}$ \lum\ (Ho \etal 1996b), and a number of other even less 
conspicuous cases have been recognized by Filippenko \& Sargent (1985).

Searching for broad H\al\ emission in nearby galaxy nuclei is a nontrivial 
business, because it entails measurement of a (generally) weak, low-contrast, 
broad emission feature superposed on a complicated stellar background.  Thus, 
the importance of careful starlight subtraction cannot be overemphasized.
Moreover, even if one were able to perfectly remove the starlight, one still 
has to contend with deblending the H\al\ + [N~II] \lamb\lamb6548, 6583 
complex.  The narrow lines in this complex are often heavily blended together, 
and rarely do the lines have simple profiles (see \S\ 6.4).  
The strategy adopted by Ho \etal (1996f) for the Palomar survey makes use of 
the line profile of the [S~II] \lamb\lamb6716, 6731 doublet to model [N~II] 
and the narrow component of H\al.  

\begin{figure}
\psfig{file=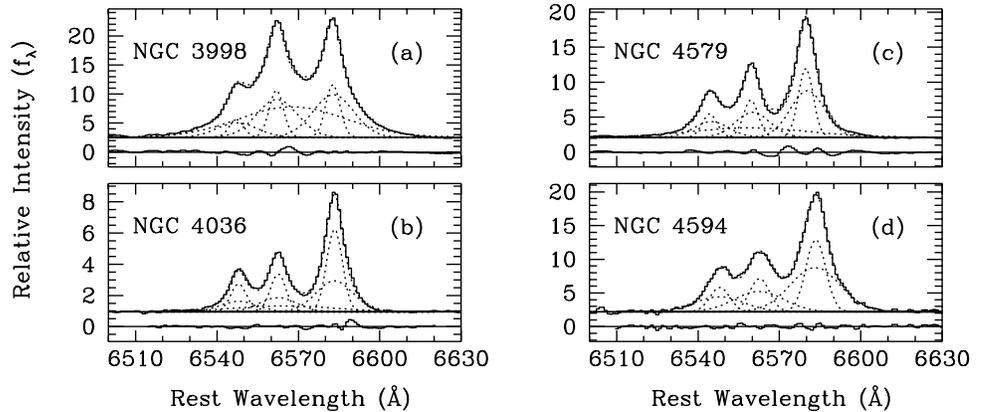,width=5.5truein,angle=-90}
\caption{Examples of LINERs with (NGC~3998, 4026, 4579) and without (NGC~4594) 
broad H\al\ emission.  [N~II] \lamb\lamb6548, 6583 and the narrow component of 
H\al\ are assumed to have the same shape as [S~II] \lamb\lamb6716, 6731, and 
the broad component of H\al\ is specified with a single Gaussian.  Residuals 
of the fit are shown on the bottom of each panel.} \label{fig10}
\end{figure}

Some examples of the line decomposition are shown in Figure~\ref{fig10}.
As was already known from previous studies (Heckman 1980b; Blackman, Wilson, 
\& Ward 1983; Keel 1983b; Filippenko \& Sargent 1985), broad H\al\ is 
unmistakably present in the LINER NGC~3998 (Fig. 10{\it a}); the line has 
FWHM $\approx$ 2150 \kms\ and FWZI \gax 5000 \kms\ (Ho \etal 1996f).  Though 
much weaker than the component in NGC~3998, a broad component of H\al\ also 
seems necessary in order to adequately model the 
H\al\ + [N~II] complex in NGC~4036 (Fig. 10{\it b}).  Broad H\al\ emission has 
long been known to exist in NGC~4579 (Stauffer 1982b; Keel 1983b; Filippenko 
\& Sargent 1985), but its strength is substantially weaker than that deduced 
by assuming that the narrow lines can be represented by single Gaussians.  The 
extended, asymmetric bases of the [N~II] lines, visible in the 
[S~II] doublet, largely account for most of the broad wings in the H\al\ + 
[N~II] blend (Fig. 10{\it c}).  Finally, I pick NGC~4594 (the Sombrero galaxy; 
Fig. 10{\it d}) to illustrate the pitfalls that can potentially afflict 
data of insufficient S/N or spectral resolution.  Judging by the 
similarity of its H\al\ + [N~II] blend to that of NGC~4579, one might be 
led to believe that NGC~4594 also has broad H\al\ emission.  However, careful
inspection of the line profiles indicates that the [S~II] lines have
large widths (FWHM $\approx$ 500 \kms) and extended wings (FWZI $\approx$ 
3000 \kms), and if one assumes that all the narrow lines have identical 
profiles, no broad H\al\ component is required to achieve a satisfactory fit 
in this object.  Note that such subtleties would easily have escaped notice 
in previous surveys.
 
Faint broad H\al\ emission has been discovered or confirmed for the first time 
in numerous nuclei.  The overall statistics of the survey can be summarized as 
follows: of the 223 emission-line nuclei classified as LINERs, transition 
objects, and Seyferts, 33 (15\%) definitely have broad H\al, and an
additional 16 (7\%) probably do.  Questionable detections were found in
another 8 objects (4\%).  Thus, approximately 20\%--25\% of all nearby
AGNs, corresponding to $\sim$10\% of all nearby, bright ($B_T\,\leq$
12.5 mag) galaxies, can be considered ``Seyfert 1'' nuclei, if one adopts the 
definition that a Seyfert 1 nucleus contains a visible BLR (see 
Table~\ref{tbl3}).  
%
%The following are from /home/lho/projects/liners/stsci_review/activity.stats
%Updated as per page 42b of notebook B16
%
\begin{table}
\caption{Percentages of Active Nuclei with Broad H\al\ Emission\tablenotemark{a},\tablenotemark{b}}\label{tbl3}
\begin{center}\scriptsize
\begin{tabular}{lcccccc}
\tableline
Hubble Type &  S & L & T & L+T & AGN & All Gals.\\
\tableline
All types   & 39  (24)                  & 24  (22)&\ 4 (3)& 16  (25)& 22  (49)& 10  (49)\\
Earlier Sbc & 40  (20)                  & 27  (22)&\ 5 (3)& 18  (25)& 24  (45)& 12  (45)\\
Later Sbc   & 27 \ (3)                  &\ 0 \ (0)&\ 0 (0)&\ 0 \ (0)&\ 9 \ (3)&\ 6 \ (3)\\
\end{tabular}
\end{center}
\tablenotetext{a}{S = Seyfert; L = LINER; T = transition object; ``L+T'' = 
LINERs + transition objects; AGN = Seyferts + LINERs + transition objects; 
``All Gals.'' = all galaxies.}
\tablenotetext{b}{The first entry is the percentage of the total in that
type, and the value given in parentheses is the actual number of objects.}
\end{table}
These numbers, of course, are merely lower limits, since undoubtedly
there must exist AGNs with even weaker broad-line emission to which we are 
insensitive. The fraction of galaxies hosting Seyfert 1 nuclei, therefore, 
is much higher than previously thought (Weedman 1977; Huchra \& Burg 1992;
Maiolino \& Rieke 1995).  Of the 33 objects with definite detections of broad
H\al, only 9 are well-known Seyfert 1 nuclei; the majority have substantially
lower H\al\ luminosities and can truly be regarded as ``dwarf'' Seyfert 1
nuclei.
 
Excluding previously classified (V\'eron-Cetty \& V\'eron 1993)
Seyfert 1 nuclei (retaining only NGC~4395; Filippenko \& Sargent 1989), the
broad H\al\ line of the remaining 40 objects has a median luminosity of
$\sim$1.2\e{39} \lum\ and FWHM = 2150 \kms\ (Ho \etal 1996f).  Five of them
have broad H\al\ luminosities as low as (1--3)\e{38} \lum, and the probable
detection in NGC~4565, if real, has a luminosity of only 8\e{37} \lum!
 
It is illuminating to consider the detection rate of broad H\al\ emission as 
a function of spectral class (Table 3).  Among objects formally classified as 
Seyferts (according to their narrow-line spectrum), approximately 40\% are 
Seyfert 1s.  The implied ratio of Seyfert 1s to Seyfert 2s (1:1.6) has 
important consequences for several models concerning the evolution and 
small-scale geometry of AGNs (e.g., Osterbrock \& Shaw 1988; Lawrence 1991), 
but such a discussion is beyond the scope of this paper and will be 
considered elsewhere.  In the present context, of greatest interest is the 
fraction of LINERs showing broad H\al\ emission.  If we first consider 
``pure'' LINERs, nearly 25\% of them have a BLR.  The detection rate among 
transition objects, however, drops drastically.  The cause for this dramatic 
change is unclear, but a likely explanation is that the broad-line component 
is simply too weak to be detected in the presence of substantial contamination 
from the H~II region component.  Supposing for the moment that the ratio of 
LINERs with and without BLRs is similar to that in Seyferts, and furthermore 
that the statistics of the presence of broad H\al\ in {\it all} LINERs (i.e., 
``pure'' LINERs + transition objects) are intrinsically the same as those of 
``pure'' LINERs, one would conclude that at least 60\% of all LINERs are 
genuine AGNs.

\section{Transition Objects: LINERs in Partial Disguise}

Following the suggestion of Ho \etal (1993), I have adopted the working 
hypothesis that the nuclei classified as transition objects represent 
composite systems consisting of contributions from both a LINER and an 
H~II region component.  Let us now consider this hypothesis in more detail.
The physical nature of this subclass of emission-line nuclei has important 
consequences for the overall demographics of LINERs and AGNs, since 
numerically these objects rival LINERs (Table 2)\footnote{Note that the 
fraction of transition objects is much higher than that given by Ho \etal 
(1993), whose estimate was based on heterogeneous data taken from the 
literature.}.

It should come as no surprise that spectra gathered from any fixed-aperture 
survey will unavoidably integrate spatially distinct regions in some objects, 
as the physical scale projected by the spectrograph aperture varies with 
distance.  Several examples of composite Seyfert/H~II region systems have 
been recognized in the literature.  In cases where the angular extent of the 
system is sufficiently large, the ``active'' component can be separated from 
the off-nuclear star-forming component (e.g., Edmunds \& Pagel 1982; 
V\'eron-Cetty \& V\'eron 1985; Shields \& Filippenko 1990).  Where only 
spatially integrated spectra are available, the composite nature has been 
identified through either decomposition of the line profiles (V\'eron \etal 
1981; Heckman \etal 1983; Kennicutt, Keel, \& Blaha 1989) or consideration 
of the line ratios (Keel 1984; Ho \etal 1993; Boer 1994).  The typical 
distances of the transition objects and LINERs in the Palomar survey, however, 
are essentially identical.  If spatial resolution is the main factor, then an 
interesting prediction, testable by high-resolution imaging, is that 
transition objects should show resolved star-forming regions surrounding a 
central LINER source (unless the star formation occurs in a compact, centrally 
located cluster).  In fact, of the galaxies in the {\it Hubble Space Telescope 
(HST)} ultraviolet imaging survey of Maoz \etal (1995; also see Maoz in
these proceedings) that show bright emission, three (NGC~4569, NGC~4736, and 
NGC~5055) have optical spectra resembling those of transition objects (Ho 
\etal 1996c), and all three exhibit resolved structure in addition to a 
central core.  Although the true nature of the ultraviolet emission can only 
be assessed through follow-up spectroscopy at comparable angular resolution, 
its morphology strongly suggests that we are witnessing star formation 
encircling an active nucleus.

Another intriguing possibility, suggested by the distribution of galaxy 
axial ratios (\S\ 6.3), is that transition objects are simply LINERs whose 
inclinations are such that circumnuclear star-forming regions happen to 
be projected along the line of sight.  Such a scenario favors a geometry in 
which the star formation in the environment of the nucleus is preferentially 
confined to a disk-like or ring-like configuration, as appears to be a common 
situation, especially for galaxies of early Hubble types (Phillips 1996).
As the emission-line strengths of LINERs in most instances are in fact weaker 
than those of giant H~II regions, an appropriate mixture of the two components 
easily accounts for the spectra of transition objects.  The average excitation 
of transition objects, as measured by [O~III]/H\bet, is lower than that of 
LINERs.  Within the framework discussed here, this finding is to be expected, 
given the high metal abundance (and hence low excitation) of nuclear H~II 
regions (Ho \etal 1996d).

Distance and orientation effects probably can account for most transition 
objects, but that cannot be the whole story, since the distributions of Hubble 
types and absolute luminosity for transition objects are actually slightly 
different compared to those of LINERs (Fig. 4), in the sense that the former 
contain some members of later morphological types.  If geometry (aperture 
and inclination effects) alone were the sole determining factor in whether 
a given nucleus is perceived as a LINER or a transition object, such 
differences would not be expected.  Could it be that some of the transition 
objects in fact do {\it not} harbor an AGN (LINER) component?  
Indeed, models attempting to explain LINERs entirely in terms of stellar 
photoionization (Filippenko \& Terlevich 1992; Shields 1992) succeed best when 
matched to objects whose [O~I] strengths (relative to H\al) are relatively 
weak\footnote{Objects originally named ``weak-[O~I] LINERs'' by Filippenko 
\& Terlevich (1992) and Ho \& Filippenko (1993) were renamed ``transition 
objects'' by Ho \etal (1993).}.  If hot stars alone contribute to the ionization
in these sources, and if the stars are not restricted to a centrally 
unresolved cluster, this alternative model can be tested through 
high-resolution imaging.

While the composite nature of transition objects demonstrates the spatial 
and temporal juxtaposition of star formation and the AGN phenomenon, it does 
{\it not} imply, much less prove, a direct causal or evolutionary connection 
between these two disparate physical processes.  Stars continuously form at 
some level in the centers of many galaxies (Ho \etal 1996d), and in early-type 
spirals, the ``hot-spot'' H~II regions can be particularly intense (e.g., 
Phillips 1996).  It has been argued (e.g., Weedman 1983) that the remnants of 
by-gone massive stars might evolve into a compact configuration at the 
nucleus, possibly in the form of a single object such as a massive black 
hole.  But until such a scenario can be proven to happen in nature, one must 
be cautious about unduly ascribing significance to a possibly fortuitous 
coexistence of two unrelated phenomena.

\section{Why are LINERs what they are?}

There is little doubt, by now, that at least some members of the LINER class
truly do belong in the AGN family.  LINERs turn out to share a surprisingly 
large number of traits found in low-luminosity Seyfert nuclei.  The global 
characteristics of their host galaxies (Hubble type, presence of a bar, 
inclination, and total luminosity), as seen at optical wavelengths at 
least, are essentially indistinguishable.  If the narrow H\al\ line can be 
regarded as an approximate gauge of the power output of the central source, it 
also appears that the luminosity of the nucleus is not a useful predictor of 
the ionization level.  Contrary to a popular misconception, not every weak 
emission-line nucleus in an early-type galaxy is a LINER; there are plenty of 
Seyfert nuclei with emission lines just as faint as, if not even fainter than, 
those seen in LINERs.  Judging by the frequency with which asymmetric 
narrow-line profiles are observed in LINERs, as well as the clear preference 
for the asymmetry to occur toward the blue half of the line center, the bulk 
velocity field of the NLRs in LINERs must have a non-negligible radial 
component, as has been known to be the case in Seyferts.  Finally, of great 
importance, the resemblance between LINERs and Seyferts has now been shown to 
extend to the presence of a BLR, one of the definitive trademarks of the AGN 
phenomenon: broad H\al\ emission is detected in roughly 25\% of LINERs.

A continuous, wide range of ionization levels clearly exists among AGNs, and 
it should be obvious that any clear-cut division of AGNs into ``high'' and 
``low'' excitation flavors is arbitrary at some level.  Instead, we should 
turn our attention to the more general question of what key parameters control 
the ionization level in AGNs.  Whenever possible, I have attempted to broach 
this issue by referencing the observed properties of LINERs with those of 
Seyferts, which, for the purposes of this discussion, are implicitly assumed 
to be a well-understood class of objects.  Rudimentary though these 
comparisons may be, some informative patterns have emerged.

As shown in \S\ 6, the NLRs in LINERs {\it differ} from those in Seyferts in 
several aspects.  The line-emitting regions tend to have lower density (at 
least for the low-density component), lower internal reddening, but larger 
line widths.  Could these indications be telling us something about 
differences in the structure of the NLR between the two types of objects?  
An additional clue, although it offers no simple explanation of the
above-mentioned observables, is furnished by the apparently higher rate
(by about a factor of two) in which density stratification, as identified
through profile variations in lines with dissimilar critical densities, is
seen in LINERs relative to Seyferts (see \S\ 6.4).

Whittle (1992a, b) finds that in Seyfert nuclei the widths of the nebular 
lines of the NLR primarily reflect the gravitational potential of the 
central region of the host galaxy.  He argues that those objects having line 
velocities that exceed the virial prediction have an additional acceleration 
mechanism, most likely in the form of radio jets emanating from the nucleus.  
Do those LINERs whose line widths are larger than those in Seyferts fall in 
this category?  One might further speculate that perhaps shocks generated as a 
consequence of this ``extra'' mechanical energy source really {\it are} 
responsible for the spectral differences between this subset of LINERs and 
Seyferts.  If shown to be true, this would be the ultimate vindication for the 
proponents of shock models (see the review by Dopita in these proceedings)!  
The conjecture that the NLRs of some LINERs experience additional acceleration 
from radio jets can be tested with appropriate radio continuum observations.  
An alternative possibility is that the line width differences actually reflect 
differences in the central mass concentration.  Future models of LINERs need 
to take all of these factors into consideration.

\section{Summary}

The main results presented in this review can be summarized as follows.

\bigskip

1) From a newly completed spectroscopic survey of nearby galaxies, 
it is confirmed that LINERs are extremely common, being present in about 
1/3 of all galaxies with $B_T\,\leq\,12.5$ mag.  If all LINERs are regarded 
as active nuclei, they constitute $>$70\% of the AGN population, and AGNs 
altogether make up nearly half of all bright galaxies.  These statistics 
should be regarded strictly as lower limits, because very faint AGNs can be 
hidden by brighter nuclear H~II regions, while others deficient in ionized 
gas may be completely invisible.

2) Approximately half of all LINERs (the so-called transition objects) show 
evidence in their integrated spectra of contamination by circumnuclear star 
formation (H~II regions).  It is argued that the majority of transition 
objects are not powered exclusively by stellar photoionization.

3) AGNs (transition objects, LINERs, and Seyferts) preferentially 
occur in early-type galaxies, mostly of Hubble types E--Sbc.  The presence of a 
bar has no visible effect on the probability of a galaxy hosting an AGN or 
on the level of activity of the AGN, when present.

4) LINERs share a number of similarities with Seyferts, but there are
several subtle differences.  The host galaxies of both classes of 
emission-line nuclei have nearly identical distributions of Hubble types, 
absolute magnitudes, and inclinations angles.  The line luminosities and the 
general properties of the bulk velocity field of their NLRs are also comparable.
However, the NLRs of LINERs differ from those of Seyferts in that the 
densities (in the low-density region) are lower, the reddenings are lower, the 
line widths are larger, and density stratification may be more common.

5) Based on the relative intensities of the narrow emission lines, at least 
10\% of all galaxies in the present survey are classified as Seyfert nuclei 
(types 1 and 2).

6) A BLR, as revealed by the presence of broad (FWHM $\approx$ 2000 \kms) 
H\al\ emission, has been detected in approximately 20\%--25\% of all nearby 
AGNs, or in $\sim$10\% of all galaxies, implying that the space-density of 
broad-lined AGNs is much higher than previously believed.  Some 25\% of LINERs 
show broad H\al\ emission.  If the ratio of LINERs with and without BLRs is 
assumed to be the same as the ratio of Seyfert 1s to Seyfert 2s (1:1.6), and if 
the low detection rate of broad H\al\ emission in transition objects can be 
attributed to observational selection effects, then at least 60\% of all 
LINERs may be genuine AGNs.

\section{Future Directions}

As much as observations at other wavelengths are opening new doors to our 
understanding of LINERs, I hope that this review has persuaded the reader that 
the more conventional technique of optical spectroscopy still has much to 
offer.  In the same spirit, I will confine my remarks on future work 
from an optical perspective.

Aside from simple statements concerning the morphological types of the host 
galaxies of LINERs and Seyferts, one can refine the treatment considerably by 
considering more quantitative measures of the bulge luminosity.  This can be 
achieved in a straightforward manner from careful bulge/disk decomposition 
(Kormendy 1977; Boroson 1981), particularly as applied to modern broad-band 
CCD images.  Sizable data bases are rapidly becoming available for many nearby 
galaxies (e.g., Prieto \etal 1992; de Jong 1996; Frei \etal 1996), but a 
concerted effort to obtain photometry for the entire Palomar sample would be 
highly desirable, and such a program is under execution.  An even more 
relevant parameter to consider is the mass of the host 
galaxy on the relevant scales; for the scale of the bulge and even smaller, 
this can be accomplished by measuring optical rotation curves in the 
traditional manner (e.g., Rubin, Whitmore, \& Ford 1988).

   With the successful identification of the large number of LINERs and
low-luminosity AGNs from ground-based surveys, we should now focus on 
identifying the key parameters in a given galaxy that regulate the level of 
activity observed.  Why do some galaxy nuclei emit such feeble power compared 
to others?  Is the mass of the central object smaller, is the accretion rate 
(relative to the Eddington rate) curtailed, or perhaps some combination of the 
two?  The amount of gaseous material required to sustain the modest observed 
luminosities of most nearby AGNs is quite small, being typically much less 
than 1 M$_\odot$ yr$^{-1}$.  Such a fueling rate conceivably can be maintained 
by mass loss from normal stars (Ho et al. 1996b), and the general interstellar 
medium from the gaseous disk must also contribute at some level; hence, it
appears that the availability of fuel should not be a major factor.  It is
possible, of course, that the accretion process itself is very inefficient,
thereby resulting in a low accretion rate.  On the other hand, if the mass of 
the central object plays the decisive role, then kinematic observations can be 
used to provide the test.  Although high-precision ground-based optical 
spectroscopy has been used to hunt for massive black holes in the centers of 
galaxies (Kormendy \& Richstone 1995), atmospheric seeing inherently limits 
the spatial resolution, and the results, although highly suggestive in several 
instances, are not conclusive.  This problem, of course, is tailor-made for 
{\it HST}, especially after the installation of the Space Telescope Imaging
Spectrograph (STIS).

By analogy with the Seyfert class, in which a continuous sequence of broad-line 
strength is seen (Osterbrock 1981), LINERs also appear to follow a similar 
sequence.  The relative visibility of the BLR in at least some Seyferts has 
been successfully interpreted in terms of orientation effects (Antonucci 1993).
It seems logical to infer that the same ``unification'' picture should extend 
to the realm of LINERs, unless some crucial element of the model (e.g., the 
presence or geometry of the obscuring molecular torus) should turn out to 
depend on the ionization state.  It would be highly worthwhile to apply the 
techniques of optical spectropolarimetry to LINERs, even though the generally 
low signal levels of these nuclei render such an experiment quite challenging.

The luminosity function of AGNs at the faint end has implications for many 
fundamental astrophysical problems.  Since LINERs constitute the bulk of the 
AGN population at low luminosities, determining the luminosity function of 
LINERs is a high priority (this work is in progress).

One obviously hopes to understand the behavior of any class of astronomical 
objects as a function of time.  However, obtaining optical spectra of a
representative sample of moderate-redshift galaxies for the identification of 
emission-line nuclei is a formidable task.  The dual need to acquire spectra 
of at least moderate quality (i.e., suitable for measuring several 
bright emission lines for classification) for a large sample of 
galaxies dictates (at the moment) the need for a multi-object spectrograph 
attached to a large (8--10 m) telescope.  As an illustration of the 
difficulty of such an enterprise, I mention two recent examples from the 
literature.  Lilly \etal (1995, and references therein) used the MOS-SIS 
spectrograph on the CFHT (3.6 m) to obtain spectra of field galaxies at 
$\langle z \rangle\, \approx\,0.6$. Despite integration times of $\sim$8 
hr, most of their spectra are inadequate for our purposes.  Even with the 
Keck 10~m telescope, moderate-quality spectra of galaxies at intermediate 
redshifts still require exposure times of $\sim$1 hr (Forbes \etal 1996).  But 
despite these difficulties, there is no obvious alternative in the near future.
Fortunately, there are several suitable samples of faint galaxies from which 
to choose.  These include the field galaxies selected from the {\it HST} 
Medium Deep Survey (Phillips \etal 1995), the $I\,\leq$ 22.5 survey of Lilly 
\etal (1995), and the Las Campanas redshift survey (Landy \etal 1996, and 
references therein).  Lastly, it should be borne in mind that even under 
conditions of optimal seeing, 1\asec\ at $z$ = 0.5 still projects to a linear 
size of $\sim$10 kpc (for $H_o$ = 75 \kms\ Mpc$^{-1}$).  At these size scales, 
the observed spectrum will include substantial contribution from the integrated 
light of the entire galaxy, and, depending on Hubble type, the signal from the 
nucleus will be severely diluted.  Thus, at intermediate redshifts, only 
the brightest nuclei will be detected, and one will not be able to quantify 
the faint end of the luminosity function.

\acknowledgments

The new results presented in this contribution were obtained in collaboration 
with Alex Filippenko and Wal Sargent and constituted a major portion of my 
Ph.D. thesis at U.~C. Berkeley.  I thank them for permission to discuss 
our work in advance of publication.  Alex Filippenko carefully read the 
manuscript and provided many helpful comments.  I am grateful to Anuradha 
Koratkar for the idea of holding this workshop, to STScI for agreeing to host 
it, and to the members of the local organizing committee for seeing it to 
fruition.  My research is currently supported by a postdoctoral fellowship 
from the Harvard-Smithsonian Center for Astrophysics.

\end{document}